\documentclass[showpacs, aps,prb,twocolumn,groupedaddress,superscriptaddress]{revtex4}
\def\ss{\vspace{0.15in}}
\def\v{\vspace{0.2in}}

\def\sv{\vspace{0.05in}}

\usepackage{graphicx}
\usepackage{color}
\usepackage{amssymb}
\usepackage{amsmath}
\usepackage{mathtools}
\usepackage{multirow}
\usepackage{anysize}

\begin{document}

% Use the \preprint command to place your local institutional report
% number in the upper righthand corner of the title page in preprint mode.
% Multiple \preprint commands are allowed.
% Use the 'preprintnumbers' class option to override journal defaults
% to display numbers if necessary
%\preprint{}

%Title of paper
%\title{Electron-Phonon Coupling in Wurtzite GaN: Intervalley Scattering Selection Rules}
\title{Electron-Phonon Coupling in Wurtzite Semiconductors: Intervalley Scattering Selection Rules for Hexagonal GaN}

% repeat the \author .. \affiliation  etc. as needed
% \email, \thanks, \homepage, \altaffiliation all apply to the current
% author. Explanatory text should go in the []'s, actual e-mail
% address or url should go in the {}'s for \email and \homepage.
% Please use the appropriate macro foreach each type of information

% \affiliation command applies to all authors since the last
% \affiliation command. The \affiliation command should follow the
% other information
% \affiliation can be followed by \email, \homepage, \thanks as well.
\author{K. M. Borysenko}\email[Email:  ]{kboryse@ncsu.edu}
\affiliation{Department of Electrical and Computer Engineering, Boston University, Boston, Massachusetts 02215}

\author{M. El-Batanouny}
\affiliation{Department of Physics, Boston University, Boston, Massachusetts 02215}

\author{E. Bellotti}
\affiliation{Department of Electrical and Computer Engineering, Boston University, Boston, Massachusetts 02215}

%\email[]{Your e-mail address}
%\homepage[]{Your web page}
%\thanks{}
%\altaffiliation{}

%Collaboration name if desired (requires use of superscriptaddress
%option in \documentclass). \noaffiliation is required (may also be
%used with the \author command).
%\collaboration can be followed by \email, \homepage, \thanks as well.
%\collaboration{}
%\noaffiliation

\date{\today}

\begin{abstract}
Selection rules are presented for electron-phonon scattering in GaN with the wurtzite crystal structure. The results are obtained for the interband scattering between the lowest conduction band ($\Gamma$-valley) and the second conduction band ($U$-valley). These selection rules are derived based on the original group-theoretical analysis of the crystal vibrations in GaN, which included detailed compatibility relations for all phonon modes.
\end{abstract}

% insert suggested PACS numbers in braces on next line \pacs{}
% insert suggested keywords - APS authors don't need to do this
%\keywords{}

%\maketitle must follow title, authors, abstract, \pacs, and \keywords
\maketitle

% body of paper here - Use proper section commands
% References should be done using the \cite, \ref, and \label commands
\section{Introduction}

Gallium nitride (GaN) and other group III nitrides have been playing
an increasingly important role in modern electronics,
optoelectronics, and photonics since 1990's; they are successfully
competing with more conventional semiconductors in various
applications. The wide band gap of these materials provides
numerous advantages for device design. In particular, it makes them
suitable for operating at high operating temperatures and electric
field strengths in high-power and microwave electronics.
Furthermore, because of the wide range of band band gap tunability,
GaN-based alloys are also desirable for use in optoelectronics form
the infrared to the blue and ultraviolet spectral range. One of the
key requirements for further successful applications of GaN and its
alloys, is better understanding of their carrier
transport properties. A stepping stone toward this goal is the
development of a comprehensive theory that describes 
carrier-phonon interaction processes.

Recently, there have been several theoretical studies of the
electron-phonon coupling in hexagonal
GaN~\cite{Yamakawa_2009_GaN_EPC_hight_field_transport_theory,Bertazzi_2009_high_field_transport_theory,Moresco_2009_avalanch_photodetectors},
where the strength of the charge carrier interaction with the
crystal lattice was estimated in the framework of the rigid
pseudo-ion model. While the results of these studies for the carrier
transport are in reasonable agreement with the available
experimental data, a more comprehensive theoretical underpinning is still required. Presently, experimental results specific to the
electron-phonon interaction processes in GaN are scarce and it is
crucial to have an independent and reliable way of validating new
theoretical predictions.

Group theory provides a unique approach to establish this
theoretical basis since it allows drawing important and far-reaching
conclusions on the characteristics of the electron-phonon
interaction processes based solely on very general symmetry
properties of the participating electron states and phonon modes.
Particularly, it allows one to establish which carrier-phonon
scattering scenarios are permitted or prohibited by symmetry.

In this work, we use the group-theoretical approach in order to
obtain detailed selection rules for the electron-phonon scattering
in the wurtzite GaN, in the same spirit as it has been done for
other materials, such as
diamond~\cite{Birman_62_selec_rules_diam_zinc_blende} and, more
relevantly, III-V semiconductors with the cubic zinc-blends crystal
structure~\cite{Birman_62_selec_rules_III_V_semicondrs}. It is
noteworthy that certain selection rules for GaN are already
available in the literature; however these works focus on optical
transitions and Raman effect~\cite{Birman_59_wurtz_opt_sel_rules,
Streitwolf_69_wurtz_selec_rules,Tronc_2001_optic_transitions,Kitaev_2001}.
In this work, on the other hand, we derive the selection rules
specifically for the electron scattering between the two lowest
conduction bands of GaN, which is a key to a better understanding of
the charge carrier transport in this semiconductor. We consider the
wurtzite GaN, a predominant crystal structure.

In order to obtain the selection rules for phonon-mediated electron
transitions, one has to (i) derive compatibility relations for all
irreducible representations (irreps) that describe the symmetry of
each phonon mode, for all high-symmetry points and lines in the (BZ); (ii) establish which irreps describe the
symmetry of the electron states before and after scattering; (iii)
based on the results of the first two steps, apply a well-known relation from the
group theory (see Eq.~(\ref{selection_rule})) in order to find which
scattering scenarios are allowed. Additionally, when one considers
the electrons in a certain valley of a conduction or valence band,
the compatibility relations for irreps that characterize the
electron states in the vicinity of the corresponding extremum point
are also necessary for a full picture. We describe these three
stages and the key results in the following sections.

\section{The Symmetry of Electron States and Compatibility Relations for Phonon Modes}

\begin{figure}[h!]
\centering
\includegraphics[bb = 32 190 392 390, width = 8cm]{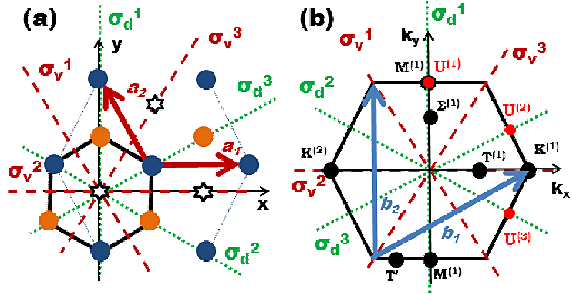}
\caption{(Color online) {\bf(a)} The projection of the wurtzite lattice on xy-plane. $\mathbf{a_1}$ and $\mathbf{a_2}$ are two independent translation vectors on xy-plane. The solid (black) line shows the borders of the symmetrized (Wigner-Seitz) unit cell. The locations of all symmetry elements on the plane are also shown: six-fold (screw) axes are marked by six-point stars; glide planes are shown as dashed (red) lines; three regular reflection planes are marked by dotted (green) lines. {\bf(b)} The Brillouin zone (BZ) of the wurtzite crystal. Two reciprocal lattice vectors in $k_x k_y$-plane ($\mathbf{b_1}$ and $\mathbf{b_2}$) are shown as blue arrows. All major points of high symmetry on this plane are marked by black dots. The projections of the three non-equivalent vectors of the star of $\mathbf{q}=U$ are also shown as smaller (red) dots. Superscripts indicate different vectors in the same star (e.g., the point $K$, whose star has two vectors, $K^{(1)}$ and $K^{(2)}$).}
\label{FIG_UnitCell_BZ}
\end{figure}

Crystal symmetry determines the classification of both electron
states and  phonon modes~\cite{Tinkham_book,MEB_book}. The general
symmetry properties of the wurtzite crystal structure have been
studied since as early as 1950's~\cite{Rashba_59_wurtzite_symmetry};
it has been found that it consists of two interpenetrating
close-packed lattices separated by $u=3/8c$ along the $z$-direction,
where $c$ is the lattice constant perpendicular to the xy-plane.
This crystal structure is described by the space group
$\mathbb{G}=C_{6v}^{4}$, which is a non-symmorphic group with 12
coset representatives with respect to the translation group
$\mathbb{T}$; half of these twelve elements are symmorphic
($\{E|0\}, 2\{C_3|0\}, 3\{\sigma_d^{(i)}|0\}$) and the other half
($\{C_2|\mathbf{\tau}\}, 2\{C_6|\mathbf{\tau}\},
3\{\sigma_v^{(i)}|\mathbf{\tau}\}$) include an improper translation
$\mathbf{\tau} = (c/2)\hat{z}$. Fig.~\ref{FIG_UnitCell_BZ}(a) shows
the projection of the wurtzite crystal lattice onto the xy-plane,
with all its symmetry elements.

%\begin{center}
%\end{center}

A simplification of the analysis is achieved in the case of the
wurtzite structure if we select the origin of coordinates on
xy-plane, at the point where it is crossed by the six-fold
rotational axis $C_6$ (as it is pictured on
Fig.~\ref{FIG_UnitCell_BZ}(a)) rather than at an atomic
site~\cite{Casella59}. The lattice basis vectors are given by
\begin{equation}
\begin{cases}
{\bf a}_1\,=\,a\,\hat{\bf x}, \\
{\bf a}_2\,=\,\frac{a}{2}\,\left(-\hat{\bf x}+\sqrt{3}\,\hat{\bf y}\right), \\
{\bf a}_3\,=\,c\hat{\bf z}
\end{cases}
\label{lattice_vectors}
\end{equation}
The atomic basis coordinates (in terms of ${\bf a}_1$-${\bf a}_3$) are
\begin{align}\boldsymbol{\rho}^\text{Ga}_1\,&=\,\Bigl(\frac{1}{3},\frac{2}{3},0\Bigr),\qquad \boldsymbol{\rho}^\text{Ga}_2\,=\,\left(\frac{2}{3},\frac{1}{3},\frac{1}{2}\right)\notag\\\boldsymbol{\rho}^\text{N}_1\,&=\,\Bigl(\frac{1}{3},\frac{2}{3},u\Bigr),\qquad \boldsymbol{\rho}^\text{N}_2\,=\,\left(\frac{2}{3},\frac{1}{3},\frac{1}{2}+u\right)\notag\end{align}
where $u=3/8\,c$. Fig.~\ref{FIG_UnitCell_BZ}(b) shows the reciprocal space with the Brillouin zone (BZ) that corresponds to this choice of coordinates. The reciprocal basis vectors are
\begin{equation}
\begin{cases}
{\bf b}_1\,=\,\frac{2\pi}{a}\left(\hat{\bf x}+\frac{1}{\sqrt{3}}\,\hat{\bf y}\right);\\
{\bf b}_2\,=\,\frac{4\pi}{\sqrt{3}a}\hat{\bf y};\\
{\bf b}_3\,=\,\frac{2\pi}{c}\hat{z}.
\end{cases}
\label{reciprocal_vectors}
\end{equation}

The $M$- and $K$-point wavevectors are $${\bf
k}_M\,=\,\frac{2\pi}{\sqrt{3}a}\hat{y};\qquad {\bf
k}_K\,=\,\frac{4\pi}{3a}\hat{x}.$$

Before we proceed with the discussion of our findings, it is
important to mention the issue of inconsistent notations that appear
in the literature for wurtzite type crystals, as well as in the
published character tables for space
groups~\cite{Rashba_59_wurtzite_symmetry, Zak_book, Slater_book,
Cracknell_book, Cardona_book, Tinkham_book}. This problem has been
discussed before~\cite{Patrick_66, Marnetto_2010_kp_wurtzite} and
the authors of the Ref.~\onlinecite{Marnetto_2010_kp_wurtzite} have
done a thorough comparison of the various existing notations. For
the sake of consistency, we will use the tables of characters from
the Ref.~\onlinecite{Zak_book} throughout this work. The reader
interested in the symmetry properties of the wurtzite crystal
structure should be aware of the discrepancies in the nomenclature
of irreps used by different authors, in order to avoid the
confusion. One important case in point is the group of the wave
vector $\mathbf{q}_\Gamma = (0,0,0)$ isomorphic to the point group
$C_{6v}$, which has six irreps $\Gamma_i$ ($i = 1, \ldots, 6$).
Comparison of the character
tables~\cite{Marnetto_2010_kp_wurtzite} shows that one of the irreps
pertinent to our analysis, namely  $\Gamma_4$, is labelled as
$\Gamma_3$ in a number of recent
works~\cite{Kitaev_2001,Nepal_2006_CB2_experiment}. At the same
time, molecular spectroscopy notations (See, for example tables in
Ref.~\onlinecite{Tinkham_book}) remain popular in the
literature~\cite{Gorczyca_95_opt_phonon_modes,Zi_96_phonons_zinc_blende_GaN,Siegle_97_Raman_experiments_GaN,Karch_98_wurtz_GaN_phonons_abinit};
in this alternative convention, $\Gamma_4$ is commonly labelled as
$B_1$. Aside from the identity representation $\Gamma_1$, the
one-dimensional irrep relevant for the hexagonal GaN is the one with
characters $(-1)$ for each non-symmorphic operation (i.e.,
$\{C_2|\mathbf{\tau}\}, 2\{C_6|\mathbf{\tau}\},
3\{\sigma_v^{(i)}|\mathbf{\tau}\}$). Lastly, we emphasize that the
difference between the character tables of $\Gamma_3$ and $\Gamma_4$
is that the characters of the glide planes $\sigma_v^{(i)}$ and the
regular planes $\sigma_d^{(i)}$ exchange signs.

%The lowest (CB1) and the second lowest (CB2) conduction bands are
%both non-degenerate; they have been identified in an early study of the electronic structure of GaN~\cite{Bloom_74_GaN_bands} as belonging to the irreps $\Gamma_1$ and $\Gamma_4$, respectively. More recent theoretical studies
%confirm these
%results~\cite{Suzuki_95_GaN_electron_bands,Marnetto_2010_kp_wurtzite,Kitaev_2001}.
%The non-degeneracy of the lowest conduction bands throughout the
%entire BZ suggests that they are derived from $s$-like states~\cite{Note_time_reversal}. One can construct
%symmetry-adapted vectors~\cite{MEB_book} from four s-orbitals, one
%for each atom in the unit cell of wurtzite GaN. The emerging
%$s$-like symmetry-adapted states can only belong to two of the four
%one-dimensional irreps of $C_{6v}$, namely $\Gamma_1$ and
%$\Gamma_4$. These results have been confirmed in several recent
%experiments~\cite{Lambrecht_95_experiment,Brazel_97_CB2_experiment,Nepal_2006_CB2_experiment}.
%The compatibility relations for the electron states near the bottoms
%of the two conduction bands are shown on
%Fig.~\ref{FIG_electron_CR}.\v

\begin{figure}[h!]
\centering
\includegraphics[bb = 61 290 258 402, width = 6cm]{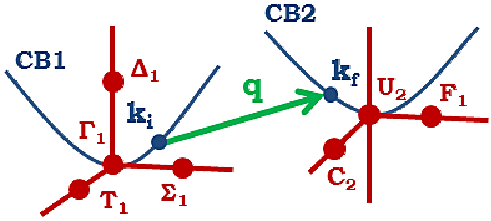}
\caption{(Color online) Compatibility relations for the electron states in two lowest conduction bands (CB1 and CB2). The green arrow signifies an arbitrary electron scattering that involves a phonon with a wave vector $\mathbf{q} = \mathbf{k_f} - \mathbf{k_i}$.}
\label{FIG_electron_CR}
\end{figure}

The s-like~\cite{Suzuki_95_GaN_electron_bands} lowest and the second lowest conduction bands (CB1 and CB2, respectively) are
 non-degenerate throughout the
entire BZ~\cite{Note_time_reversal}. They have been identified in an early study of the electronic structure of GaN~\cite{Bloom_74_GaN_bands} as belonging to the irreps $\Gamma_1$ (CB1) and $\Gamma_4$ (CB2). More recent theoretical~\cite{Suzuki_95_GaN_electron_bands,Marnetto_2010_kp_wurtzite,Kitaev_2001} and experimental~\cite{Lambrecht_95_experiment,Brazel_97_CB2_experiment,Nepal_2006_CB2_experiment} studies
confirm these
results. The electronic state at the minimum of the $U$-valley has a symmetry $U_2$, compatible with $\Gamma_4$~\cite{Bloom_74_GaN_bands,Lambrecht_95_experiment}. The compatibility relations for the electron states near the bottoms
of the two conduction bands are shown on
Fig.~\ref{FIG_electron_CR}.

%Indeed, both lowest conduction bands are non-degenerate, which suggests their $s$-like nature. One can construct symmetry-adapted vectors~\cite{MEB_book} from four s-orbitals, one for each atom in the unit cell of wurtzite GaN. The obtained $s$-like states demonstrate the symmetry that is compatible with two of the four one-dimensional irreps of $C_{6v}$, namely $\Gamma_1$ and $\Gamma_4$. These results have been confirmed again for the two lowest conduction bands in several more recent experiments~\cite{Lambrecht_95_experiment,Brazel_97_CB2_experiment,Nepal_2006_CB2_experiment}.

Having established the symmetry of the relevant electron states in hexagonal GaN, it is now necessary to comprehensively describe the symmetry of atomic thermal vibrations in wurtzite. This implies deriving compatibility relations between all  points of high symmetry in the BZ. Generally, these relations allow for several equally valid solutions. For example, as $\mathbf{q_{\Sigma}} \rightarrow \mathbf{q_{M}}$, the compatibility relations are
\begin{equation*}
\Sigma_1 \rightarrow M_1,M_4; \;
\Sigma_2 \rightarrow M_2,M_3.
\end{equation*}
Therefore, if a certain phonon branch's symmetry is described by $\Sigma_1$ along the $\Sigma$-direction, one has to still establish which of the two compatible irreps describes that branch at the $M$-point. To eliminate such ambiguity, we have obtained symmetry-adapted vectors for atomic displacements in the wurtzite crystal at the $\Gamma$-point (The procedure for obtaining these vectors, based on symmetry projection operators, is described in detail in Chapter~6 of Ref.~\onlinecite{MEB_book}). The phonon polarization vectors at a given $\mathbf{q}$-point of the BZ can be readily obtained by constructing normalized linear combinations of the symmetry-adapted vectors. The resulting vectors for the $\Gamma$-point can be found in Appendix~A.

%In order to do that, one has to consider change of symmetry that occurs between any two adjacent $q$-points $\mathbf{q_A}$ and $\mathbf{q_B}$, i.e. points
%
% apply a well-known relation of the group theory to calculate the frequencies $a_i$ of each irrep in a given point $\mathbf{q_B}$
%\begin{equation}
%a_i = \frac{1}{h}\sum_{g \in (G^{\mathbf{q_A}}\bigcap G^{\mathbf{q_B}})} \left[\chi^{\mathbf{q_B}}_{i}(g)\right]^{*} \chi^{\mathbf{q_A}}_{j}(g)
%\end{equation}
%for two adjacent points $\mathbf{q_A}$ and $\mathbf{q_B}$ in the BZ with different symmetry, where $h$ is a number of elements in the intersection of the groups of these wave vectors $(G^{\mathbf{q_A}}\bigcap G^{\mathbf{q_B}})$; $\chi^{\mathbf{q_A}}_{j}(g)$ and $\chi^{\mathbf{q_B}}_{i}(g)$ are the characters of the $j$th irrep of $G^{\mathbf{q_A}}$ and $i$th irrep of $G^{\mathbf{q_B}}$, respectively, for a given group element $g = \left\{R|\tau(R)\right\}$.
%

Three acoustic phonon modes are easily identified among the obtained vectors as:
\begin{equation}
\begin{cases}
|\Gamma^{(ac)},x\rangle = |5,1\rangle, \\
|\Gamma^{(ac)},y\rangle = |5,3\rangle, \\
|\Gamma^{(ac)},z\rangle = |1,1\rangle,
\end{cases}
\label{acoustic_modes}
\end{equation}
(See Eqs.~(\ref{polar_vects_G1_1})-(\ref{polar_vects_G5})), in agreement with Ref.~\onlinecite{Birman_CdS} which assigns $\Gamma_1 \oplus \Gamma_5$ to three acoustic phonon modes at the point $\Gamma$. The remaining vectors in Eqs.~(\ref{polar_vects_G1_1})-(\ref{polar_vects_G6}) also agree with atomic displacement patterns for optical modes published earlier~\cite{Gorczyca_95_opt_phonon_modes}. By applying this procedure to all pertinent points of high symmetry we generate all polarization vectors, as well as compatible irrep frequencies $a_i$. The following relations show how the reducible representation $D$ that describes \emph{all} atomic displacements in the wurtzite crystal, splits into compatible irreps in all major points of high symmetry on $k_xk_y$-plane in the BZ:
\begin{subequations}
\begin{eqnarray}
D^{\Gamma} &=& 2\Gamma_1 \oplus 2\Gamma_4 \oplus 2\Gamma_5 \oplus 2\Gamma_6, \label{Irrep_freqs_in_Gamma}\\
D^{T} &=& 6T_1 \oplus 6T_2, \\
D^{K} &=& 2K_1 \oplus 2K_2 \oplus 4K_3, \\
D^{\Sigma} &=& 8\Sigma_1 \oplus 4\Sigma_2, \\
D^{M} &=& 4M_1 \oplus 2M_2 \oplus 2M_3 \oplus 4M_4, \\
D^{T^{\prime}} &=& 6T^{\prime}_1 \oplus 6T^{\prime}_2.
\label{Irrep_freqs_in_T}
\end{eqnarray}
\end{subequations}

\begin{figure}[h!]
\centering
\includegraphics[bb = 34 230 307 420, width = 8cm]{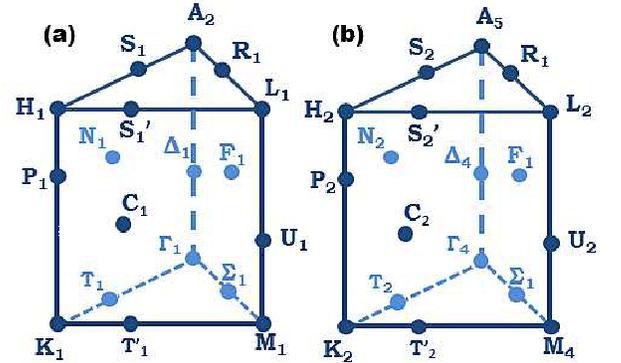}
\caption{Compatibility relations for the phonon modes $\Gamma_1$ (a) and $\Gamma_4$ (b) in the irreducible wedge of the BZ. All irrep labels correspond to the convention used in the character tables in Ref.~\onlinecite{Zak_book}. An arbitrary point in the bulk of the irreducible wedge $\Theta_1^{\prime}$, is not shown as it does not have any symmetry operations associated with it and is irrelevant to the selection rules discussed in this work.}
\label{FIG_CR_G1_G4}
\end{figure}

\begin{figure}[h!]
\centering
\includegraphics[bb = 60 150 303 383, width = 7cm]{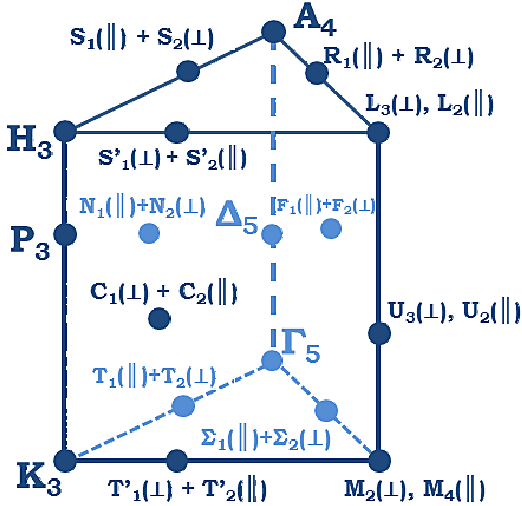}
\caption{Compatibility relations for the phonon mode $\Gamma_5$ in the irreducible wedge of the BZ. A symbol $\parallel$ ($\perp$) next to the name of an irrep signifies that the latter describes a mode at a given point $\mathbf{q}$, with the the polarization vector parallel (perpendicular) to the reflection plane in which the vector $\mathbf{q}$ lies. Symbols with a larger font (e.g., $\Delta_5$) mark the points inside the BZ where two phonon modes are allowed to be degenerate by symmetry and thus, are described by two-dimensional irreps. Note that the plus sign is used whenever a certain point is adjacent to one where two modes are degenerate by symmetry; therefore, it emphasizes that the degeneracy is lifted due to a reduced symmetry.\v}
\label{FIG_CR_G5}
\end{figure}

\begin{figure}[h!]
\centering
\includegraphics[bb = 58 150 317 402, width = 7cm]{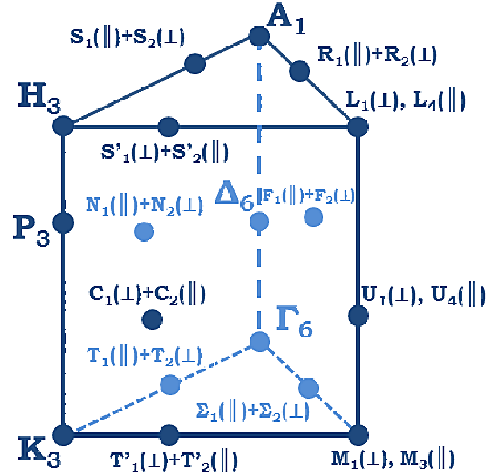}
\caption{Compatibility relations for the phonon mode $\Gamma_6$ in the irreducible wedge of the BZ. See also the caption to Fig.~\ref{FIG_CR_G5}.}
\label{FIG_CR_G6}
\end{figure}

Since no group of a wave vector in the BZ of the wurtzite crystal contains operations that affect $z$-component of $\mathbf{q}$,~\cite{Rashba_59_wurtzite_symmetry} the phonon modes with the same $(q_x,q_y)$ have the same symmetry-adapted vectors associated with them. In other words, polarization vectors $\mathbf{e}_{\mu}(\mathbf{q})$ that determine atomic displacements for each mode $\mu = 1, \ldots, 12$, are independent of $q_z$. The summary of the obtained compatibility relations is presented on Figs.~(\ref{FIG_CR_G1_G4}-\ref{FIG_CR_G6}). To the best of our knowledge, the only work that has previously shown detailed compatibility relations was Ref.~\onlinecite{Birman_CdS}, which studied the phonon spectrum of CdS with the wurtzite crystal structure. These results should be completely applicable to any crystal with the wurtzite structure. However, a careful analysis reveals several inconsistencies in the compatibility relations of Ref.~\onlinecite{Birman_CdS}. Particularly, they do not agree with the results of Eqs.~(\ref{Irrep_freqs_in_Gamma}-\ref{Irrep_freqs_in_T}). The recalculated compatibility relations shown on Figs.~(\ref{FIG_CR_G1_G4}-\ref{FIG_CR_G6}) resolve this issue.

\section{Selection Rules}

A selection rule in the case of electron-phonon interaction provides an answer to a question whether the Kronecker product of three irreps $\left(\left[D_\gamma^{\mathbf{k_f}}\right]^{*}\otimes D_\beta^{\mathbf{q}}\otimes D_\alpha^{\mathbf{k_i}}\right)$ contains the identity representation, where $D_\alpha^{\mathbf{k_i}}, D_\gamma^{\mathbf{k_f}}$, and $D_\beta^{\mathbf{q}}$ are the representations of, respectively, the group of the wave vector of the electron  $G^{\mathbf{k_i}}$ (initial state, i.e., before scattering), $G^{\mathbf{k_f}}$ (final state), and the phonon $G^{\mathbf{q}}$ involved in scattering ($\mathbf{q} = \mathbf{k_f} - \mathbf{k_i}$, by momentum conservation). If the answer is positive, then such scattering process is allowed by symmetry. Formally, one has to find the frequency of the identity representation for each combination of three irreps $D_\alpha^{\mathbf{k_i}}, D_\gamma^{\mathbf{k_f}}$, and $D_\beta^{\mathbf{q}}$, according to the well-known relation~\cite{Tinkham_book,MEB_book}
\begin{equation}
a_1 = \frac{1}{h}\sum_{g \in \left(G^{\mathbf{k_f}}\bigcap G^{\mathbf{q}}\bigcap G^{\mathbf{k_i}}\right)} \left[\chi^{\mathbf{k_f}}_{\gamma}(g)\right]^{*} \chi^{\mathbf{q}}_{\beta}(g) \chi^{\mathbf{k_i}}_{\alpha}(g),
\label{selection_rule}
\end{equation}
where $h$ is the number of elements in the group $\left(G^{\mathbf{k_f}}\bigcap G^{\mathbf{q}}\bigcap G^{\mathbf{k_i}}\right)$, and $\chi^{\mathbf{k}}_{\alpha} = Tr\left[D_\alpha^{\mathbf{k}}\right]$ is the character of the $\alpha$th irrep of the wave group of $\mathbf{k}$, $D_\alpha^{\mathbf{k}}$. If $a_1 = 0$ for chosen $D_\alpha^{\mathbf{k_i}}, D_\gamma^{\mathbf{k_f}}$, and $D_\beta^{\mathbf{q}}$, then such scattering scenario is prohibited by symmetry.

A summary of all non-trivial selection rules for the interband scattering (CB1-CB2) between $\Gamma$- and $U$-valleys in hexagonal GaN is presented in Table~\ref{T_all_SR}. Three sections of the table show scattering scenarios for three high-symmetry points/directions in $\Gamma$-valley of CB1: $\Gamma$, $\Sigma$, and $T$. Each row in a table represents a specific scattering process, when the initial and final electron states are defined. Note that the selection rules for the initial electron state with the wave vector $\mathbf{k}=(k_x,k_y,k_z)$ are exactly the same as for the case when this vector is projected on $xy$-plane: $\mathbf{k^\prime}=(k_x,k_y,0)$. So for example, electron scattering from a $\Delta$-point follows the same rules as scattering from $\mathbf{k}=\boldsymbol{\Gamma}$. Ultimately, this is due to the property mentioned earlier, that all symmetry operations of the wurtzite type crystal leave z-component of the wave vector unchanged.

%$\Gamma_1 \left(\Delta_1 \right)$, $\Sigma_1 \left(F_1 \right)$, and $T_1 \left(N_1  \right)$, respectively.  %First column of each table contains irreps for the electron states in specific points of $U$-valley (the final state). Second column shows one or more irreps permitted by the selection rule, for the phonon modes that satisfy the momentum conservation for the given initial and final states of the electron. The next four columns of the table contain the actual selection rules for each of the four types of phonon modes labelled by their irreps at the $\Gamma$-point: $\Gamma_1,\Gamma_4,\Gamma_5,\Gamma_6$. "Yes" ("No") in any of these four columns means that all (none) of the phonon modes of this type are allowed to participate in a specified scattering scenario. If only certain polarization vectors of a given mode are allowed, they are shown explicitly [See Eqs.~(\ref{polar_vects_G1_1}-\ref{polar_vects_G6}) for a full list of vectors]. Finally, in some cases it is necessary to choose also a specific $\mathbf{q}$-vector of a star as that, generally, determines which symmetry operations will be included in the group $\left(G^{\mathbf{k_f}}\bigcap G^{\mathbf{q}}\bigcap G^{\mathbf{k_i}}\right)$. In such case, an additional superscript is added to the irrep notation. For instance, the index $"(3)"$ in $U_2^{(3)}$ (Table \ref{T_all_SR}) emphasizes that only one of the three non-equivalent $U$-points in the star abides by the shown selection rules (All three points in the star $\mathbf{k}=\boldsymbol{U}$ are shown on Fig.~\ref{FIG_UnitCell_BZ}).
Finally, one can think of more scattering scenarios  between $\Gamma$- and $U$-valleys than what is presented in the Table~\ref{T_all_SR}, due to all possible combinations of vectors in the stars of $\mathbf{q}$ and $\mathbf{k_f}$ (we can assume that $\mathbf{k_i}$ is fixed). However, analysis reveals that many of these combinations correspond to the same (lowest symmetry) case when the intersection of the groups $\left(G^{\mathbf{k_f}}\bigcap G^{\mathbf{q}}\bigcap G^{\mathbf{k_i}}\right)$ contains only one element, $\left\{E|0\right\}$ and thus, according to Eq.~(\ref{selection_rule}) all phonon modes in wurtzite GaN can participate in such scattering process. All these cases are excluded from the Table~\ref{T_all_SR} as trivial, with the exception of the last  row of the case (II). The latter illustrates the situation when three groups of the wave vectors $G^{\mathbf{k_f}}, G^{\mathbf{q}}$, and $G^{\mathbf{k_i}}$, while generally possessing certain symmetries individually ($\{\sigma_d^1|0\} \in G^{\mathbf{k_i}}$, $ \{\sigma_v^2|\tau\} \in G^{\mathbf{k_f}}$, $\{\sigma_v^3|\tau\} \in G^{\mathbf{q}}$), have no common symmetry operations except the identity $\left\{E|0\right\}$. \ss
\begin{table}
\begin{center}\caption{\label{T_all_SR} A summary of the selection rules for the case when the electron is being scattered by a phonon from the $\Gamma$-valley in CB1, to the $U$-valley in CB2. Three specific points in CB1 are considered: (I) $\Gamma$-point (or a $\Delta$-point); (II) $\Sigma$-point (or an $F$-point) ; (III) $T$-point (or an $N$-point). The subscripts identify the irreps that describe the symmetry of the electron states and the phonon mode involved in the scattering process. The superscript (e.g., in $U_2^{(3)}$) specifies which vector in the star $S_\mathbf{q}$ is being considered (See Fig.~\ref{FIG_UnitCell_BZ}(b)). The notation $\left\{ U_2  \right\}$ is used for the minimum of the $U$-valley together with its neighborhood. The first column shows the irrep that describes the final electron state. The second column shows which phonon irrep $D_\beta^{\mathbf{q}}$ (with $\mathbf{q} = \mathbf{k_f} - \mathbf{k_i}$) is allowed by selection rules for the specified initial and final electron states. The last column shows which phonon modes of wurtzite GaN are compatible with a given $D_\beta^{\mathbf{q}}$. Since only one final state ($U_2^{(1)}$ (or $C_2^{(1)}$, which leads to the same selection rules)) is considered in case (III), the first column in this section is left empty.}\sv
\begin{tabular}{|c|c||c|}
\hline
\hline
\multicolumn{3}{|c|}{ (I) \hspace{0.2in} $ D_\alpha^{\mathbf{k_i}} = \Gamma_1   \hspace{0.2in} \stackrel{D_\beta^{\mathbf{q}}}{\longrightarrow} \hspace{0.2in} D_\gamma^{\mathbf{k_f}} = \left\{ U_2  \right\}$ } \\[3pt]
\hline
%
%$D_\alpha^{\mathbf{k_i}}$ &
$D_\gamma^{\mathbf{k_f}}$ & $D_\beta^{\mathbf{q}}$  & Allowed modes \\[3pt]
\hline
\hline
%
%   $\Gamma_1 \left(\Delta_1 \right)$  &
   $U_2$  & $U_2$   & $|4,\left\{1,2\right\}\rangle, |5,\left\{3,4\right\}\rangle$\\[3pt]
%
%   $\Gamma_1 \left(\Delta_1 \right)$  &
   $F_1$  & $F_1$ & $|1,\left\{1,2\right\}\rangle,|4,\left\{1,2\right\}\rangle, |5,\left\{3,4\right\}\rangle, |6,\left\{3,4\right\}\rangle $  \\[3pt]
%   $\Gamma_1 \left(\Delta_1 \right)$  &
   $C_2$  & $C_2$  & $|4,\left\{1,2\right\}\rangle,|5,\left\{3,4\right\}\rangle,|6,\left\{1,2\right\}\rangle $  \\
\hline
\hline
\hline
\multicolumn{3}{|c|}{ (II) \hspace{0.2in} $  D_\alpha^{\mathbf{k_i}} = \Sigma_1^{(1)}  \hspace{0.1in} \stackrel{D_\beta^{\mathbf{q}}}{\longrightarrow} \hspace{0.1in} D_\gamma^{\mathbf{k_f}} = \left\{ U_2^{(1)}  \right\}$ } \\[3pt]
\hline
% This line is the same for all tables:
%$D_\alpha^{\mathbf{k_i}}$ &
$D_\gamma^{\mathbf{k_f}}$ & $D_\beta^{\mathbf{q}}$  & Allowed modes \\[3pt]
\hline
\hline
%
%$\Sigma_1^{(1)} \left(F_1^{(1)} \right)$ &
$U_2$  & $F_1$   & $|1,\left\{1,2\right\}\rangle, |4,\left\{1,2\right\}\rangle, |5,\left\{3,4\right\}\rangle,|6,\left\{3,4\right\}\rangle $  \\
%
%$\Sigma_1^{(1)} \left(F_1^{(1)} \right)$ &
$F_1$  & $F_1$   & $|1,\left\{1,2\right\}\rangle, |4,\left\{1,2\right\}\rangle, |5,\left\{3,4\right\}\rangle,|6,\left\{3,4\right\}\rangle $  \\
%$\Sigma_1^{(1)} \left(F_1^{(1)} \right)$ &
%$C_2$  & $\Theta_1^{\prime}$  & All modes are allowed  \\
%$\Sigma_1^{(1)} \left(F_1^{(1)} \right)$ &
$C_2$  & $N_{1,2}$            & All modes are allowed \\
\hline
\hline
\hline
\multicolumn{3}{|c|}{ (III) \hspace{0.2in} $  D_\alpha^{\mathbf{k_i}} = T_1^{(1)}  \hspace{0.1in} \stackrel{D_\beta^{\mathbf{q}}}{\longrightarrow} \hspace{0.1in} D_\gamma^{\mathbf{k_f}} = U_2^{(1)} (C_2^{(1)})$ } \\[3pt]
\hline
%$D_\alpha^{\mathbf{k_i}}$ &
$D_\gamma^{\mathbf{k_f}}$ & $D_\beta^{\mathbf{q}}$  & Allowed modes \\[3pt]
\hline
\hline
%$T_1^{(1)} \left(N_1^{(1)}  \right)$ &
  & $C_2$   & $|4,\left\{1,2\right\}\rangle, |5,\left\{3,4\right\}\rangle, |6,\left\{1,2\right\}\rangle$  \\
%
%$T_1^{(1)} \left(N_1^{(1)}  \right)$ &
 & $P_{2,3}$   & $|4,\left\{1,2\right\}\rangle, |5,\left\{1,\ldots,4\right\}\rangle, |6,\left\{1,\ldots,4\right\}\rangle$ \\
\hline
\hline
\end{tabular}
\end{center}\end{table}
\section{Conclusions}
In this work, we have derived the selection rules for electron scattering between the two conduction bands in the wurtzite GaN. This analysis is based on the detailed compatibility relations for all phonon modes, which revealed some inconsistencies in the earlier work~\cite{Birman_CdS}. Due to a relatively low symmetry of the wurtzite crystal structure, a majority of the phonon modes are allowed to participate in the electron scattering between the valleys at $\mathbf{q}=\boldsymbol{\Gamma}$ (CB1) and $\mathbf{q}=\boldsymbol{U}$ (CB2), with the exception of the $z$-polarized $\Gamma_1$-modes (both acoustic and optical), which have shown a number of restrictions along high-symmetry directions of the BZ. Parallel-vs-perpendicular nature of the modes (i.e. whether a given polarization vector is parallel or perpendicular to a reflection plane that contains the vector $\mathbf{q}$) also imposes certain restrictions in case of the phonons described by the irreps $\Gamma_5$ and $\Gamma_6$. These results provide a reliable sanity check for future theoretical studies that involve electron-phonon coupling in hexagonal GaN, as they are expected to dictate the overall behavior of the corresponding matrix elements.
\section{Acknowledgment}
KMB and EB gratefully acknowledge financial support from
the U. S. Army Research Laboratory through the Collaborative
Research Alliance (CRA) for MultiScale multidisciplinary Modeling of
Electronic materials (MSME).
\appendix
\section{Polarization vectors of the phonon modes in the wurtzite crystal}
The following is a list of the twelve phonon polarization vectors at the point $\Gamma$, divided among four different irreps (See Eq.~(\ref{Irrep_freqs_in_Gamma})):
\begin{subequations}
\begin{eqnarray}
2\Gamma_1: \nonumber \\
|1,1\rangle & = & \frac{1}{2}\left\{\hat{z},\hat{z},\hat{z},\hat{z} \right\}, \hspace{0.2in} \label{polar_vects_G1_1} \\
%|1,1\rangle & = & \frac{1}{2}\left\{1,1,1,1 \right\}_z, \label{polar_vects_G1_1} \\
|1,2\rangle & = & \frac{1}{2}\left\{\hat{z},\hat{z},-\hat{z},-\hat{z}\right\},\hspace{0.35in}
\label{polar_vects_G1_2}
\end{eqnarray}
\end{subequations}
\begin{subequations}
\begin{eqnarray}
2\Gamma_4: \nonumber \\
|4,1\rangle & = & \frac{1}{2}\left\{-\hat{z},\hat{z},-\hat{z},\hat{z} \right\}, \hspace{0.35in} \\
|4,2\rangle & = & \frac{1}{2}\left\{-\hat{z},\hat{z},\hat{z},-\hat{z} \right\}, \hspace{0.35in}
\label{polar_vects_G4}
\end{eqnarray}
\end{subequations}
\begin{subequations}
\begin{eqnarray}
2\Gamma_5: \nonumber \\
|5,1\rangle & = & \frac{1}{2}\left\{\hat{x},\hat{x},\hat{x},\hat{x} \right\}, \hspace{0.3in} \\
|5,2\rangle & = & \frac{1}{2}\left\{ \hat{x},\hat{x},-\hat{x},-\hat{x} \right\}, \hspace{0.3in} \\
|5,3\rangle & = & \frac{1}{2}\left\{\hat{y},\hat{y},\hat{y},\hat{y} \right\}, \hspace{0.35in}\\
|5,4\rangle & = & \frac{1}{2}\left\{\hat{y},\hat{y},-\hat{y},-\hat{y} \right\}, \hspace{0.35in}
\label{polar_vects_G5}
\end{eqnarray}
\end{subequations}
\begin{subequations}
\begin{eqnarray}
2\Gamma_6: \nonumber \\
|6,1\rangle & = & \frac{1}{2}\left\{-\hat{x},\hat{x},-\hat{x},\hat{x} \right\}, \hspace{0.35in} \\
|6,2\rangle & = & \frac{1}{2}\left\{-\hat{x},\hat{x},\hat{x},-\hat{x} \right\}, \hspace{0.35in}\\
|6,3\rangle & = & \frac{1}{2}\left\{-\hat{y},\hat{y},-\hat{y},\hat{y} \right\}, \hspace{0.35in} \\
|6,4\rangle & = & \frac{1}{2}\left\{-\hat{y},\hat{y},\hat{y},-\hat{y} \right\}. \hspace{0.35in}
\label{polar_vects_G6}
\end{eqnarray}
\end{subequations}
These vectors are 12-dimensional (four atoms in the wurtzite unit cell, times three Cartesian coordinates). We, however, use a more compact, four-component notation, where the displacement of each of four atoms of the basis is described by an appropriate unit vector. So for instance, an atomic displacement along x-direction is described by the vector $\hat{x} = (1,0,0)$.
These twelve vectors represent all possible atomic displacement patterns in the wurtzite crystal. They form a basis that spans the corresponding Hilbert space, so that any phonon mode at a point $\mathbf{q}\ne \boldsymbol{\Gamma}$ of the BZ can also be represented in terms of these vectors. In other words, Eqs.~(\ref{polar_vects_G1_1}-\ref{polar_vects_G6})
define basis functions for \emph{all} compatible irreps at different points in the BZ of the wurtzite crystal.
\begin{table}[h]
\begin{center}\caption{\label{TA} A list of high-symmetry points on $q_xq_y$-plane of the BZ of wurtzite (first column), with associated irreps (second column; see also Figs.~(\ref{FIG_CR_G1_G4}-\ref{FIG_CR_G6})), and corresponding polarization vectors (third column). The superscripts in the names of the high-symmetry points indicate the choice of the vector in the star (See Fig.~\ref{FIG_UnitCell_BZ}). If several eigenvectors $|\alpha, i \rangle,.., |\alpha, j \rangle $ of an irrep $\Gamma_\alpha$ serve as basis functions for an irrep specified in the second column, their full list is shown in curly braces: $|\alpha, \left\{ i,..,j \right\} \rangle $. } \sv
\begin{tabular}{|l|l|l|}
\hline
\hline
\multirow{2}{*}{$\Sigma^{(1)}$} & $\Sigma_1$ & $|1,\left\{ 1,2 \right\}\rangle,|4,\left\{ 1,2 \right\}\rangle,  |5,\left\{ 3,4 \right\}\rangle, |6,\left\{ 3,4 \right\}\rangle$  \\[3pt]
 & $\Sigma_2$ & $|5,\left\{ 1,2 \right\}\rangle, |6,\left\{ 1,2 \right\}\rangle $ \\[3pt] %\hline
\hline
\multirow{4}{*}{$M^{(1)}$} & $M_1$ & $|1,\left\{ 1,2 \right\}\rangle,|6,\left\{ 3,4 \right\}\rangle$ \\[3pt]
 & $M_2$ & $|5,\left\{ 1,2 \right\}\rangle$ \\ [3pt]
 & $M_3$ & $|6,\left\{ 1,2 \right\}\rangle$ \\ [3pt]
 & $M_4$ & $|4,\left\{ 1,2 \right\}\rangle, |5,\left\{ 3,4 \right\}\rangle$ \\ [3pt]%\hline
\hline
\multirow{2}{*}{$T^{(1)}$} & $T_1$ & $|1,\left\{ 1,2 \right\}\rangle, |5,\left\{ 1,2 \right\}\rangle, |6,\left\{ 3,4 \right\}\rangle$ \\[3pt]
 & $T_2$ & $|4,\left\{ 1,2 \right\}\rangle, |5,\left\{ 3,4 \right\}\rangle, |6,\left\{ 1,2 \right\}\rangle$ \\ [3pt]%\hline
\hline
\multirow{3}{*}{$K^{(1)}$} & $K_1$ & $|1,\left\{ 1,2 \right\}\rangle$ \\[3pt]
 & $K_2$ & $|4,\left\{ 1,2 \right\}\rangle$ \\ [3pt]
 & $K_3$ & $|5,\left\{1,\ldots,4\right\}\rangle, |6,\left\{1,\ldots,4\right\}\rangle$ \\ [3pt] \hline
\hline

\end{tabular}
\end{center}\end{table}
Table~\ref{TA} shows a list of all compatible irreps (See Figs.~(\ref{FIG_CR_G1_G4}-\ref{FIG_CR_G6})) for high-symmetry $\mathbf{q}$-points in $q_xq_y$-plane, with corresponding polarization vectors. This table, together with the vectors (\ref{polar_vects_G1_1}-\ref{polar_vects_G6}) and the frequencies of irreps at different points of the BZ (See Eqs.~(\ref{Irrep_freqs_in_Gamma}-\ref{Irrep_freqs_in_T})), provides a comprehensive picture of the symmetry of thermal atomic vibrations in the wurtzite GaN. Note that it is sufficient to consider only points in $q_xq_y$-plane, since all modes on the same phonon branch and with the same $(q_x,q_y)$ are characterized by the same polarization vectors (See the brief discussion at the end of Section II, in regards to Ref.~\onlinecite{Rashba_59_wurtzite_symmetry}).


\begin{thebibliography}{9}
\bibitem{Note_time_reversal} Their
degeneracy on the surface of the BZ that contains the triangle
$A-H-L$ can be explained by symmetry with respect to time reversal,
in full analogy with what is observed in the phonon spectrum of the
wurtzite crystal~\cite{Patrick_66}
\end{thebibliography}
\end{document}